# Ultrafast transient generation of spin density wave order in the normal state of BaFe$_2$As$_2$ driven by coherent lattice vibrations

K. W. Kim[1,2,3], A. Pashkin[1], H. Schäfer[1], M. Beyer[1], M. Porer[1,4], T. Wolf[5], C. Bernhard[2], J. Demsar[1,6], R. Huber[1,4], and A. Leitenstorfer[1]

[1]Department of Physics and Center for Applied Photonics, University of Konstanz, D-78457 Konstanz, Germany.

[2]University of Fribourg, Department of Physics and Fribourg Center for Nanomaterials, Chemin du Musée 3, CH-1700 Fribourg, Switzerland.

[3]Department of Physics, Chungbuk National University, Cheongju 361-763, Korea.

[4]Department of Physics, University of Regensburg, D-93053 Regensburg, Germany.

[5]Karlsruhe Institute of Technology, Institute for Solid-State Physics, D-76021 Karlsruhe, Germany.

[6]Complex Matter Department, Jozef Stefan Institute, SI-1000 Ljubljana, Slovenia.

**The interplay among charge, spin and lattice degrees of freedom in solids gives rise to intriguing macroscopic quantum phenomena such as colossal magneto-resistance, multiferroicity, and high-temperature superconductivity[1-3]. Strong coupling or competition between various orders in these systems presents the key to manipulate their functional properties via external perturbations such as electric and magnetic fields[2] or pressure[3]. Ultrashort and intense optical pulses have emerged as an interesting tool to investigate elementary dynamics and control material properties by melting an existing order[4-6]. In this work, we employ few-cycle multi-THz pulses to resonantly probe the evolution of the spin density wave (SDW) gap of the pnictide compound BaFe$_2$As$_2$ following excitation with a femtosecond optical pulse. When**

**starting in the low-temperature ground state, optical excitation results in a melting of the SDW order, followed by ultrafast recovery. In contrast, the SDW gap is *induced* when we excite the normal state above the transition temperature. Very surprisingly, the transient ordering quasi-adiabatically follows a coherent lattice oscillation at a frequency as high as 5.5 THz. Our results attest to a pronounced spin-phonon coupling in pnictides that supports rapid development of a macroscopic order upon small vibrational displacement even without breaking the symmetry of the crystal.**

Within the past few years, the pnictides have been found as a new class of unconventional superconductors[3,7-13]. Like in cuprates, high-temperature superconductivity is implemented by doping of an antiferromagnetic pnictide parent compound with carriers[1,9,10]. Such a phase diagram with superconductivity emerging from a magnetic ground state has been observed in many other unconventional superconductors[14]. Therefore, the underlying magnetic order has been at the heart of the discussion concerning the mechanism of high-temperature superconductivity in general[10-13]. $BaFe_{2-x}Co_xAs_2$ is one of the most studied Fe-based superconductors due to the availability of large and high-quality single crystals. An antiferromagnetic SDW represents the ground state in the undoped compound $BaFe_2As_2$. Electron doping with Co gradually suppresses the SDW transition temperature $T_{SDW}$ and superconductivity emerges[11]. Interestingly, in the underdoped region the two phases are found to coexist on a length scale smaller than a few nanometers[12,13]. In addition, $BaFe_{2-x}Co_xAs_2$ undergoes a structural phase transition from tetragonal to orthorhombic at or slightly above $T_{SDW}$.

Another structural phase transition back to the tetragonal symmetry was reported in the superconducting state in samples close to optimal doping[11]. Experiments under external pressure have confirmed the intimate connection between the lattice structure and electronic ordering, with increasing pressure suppressing the SDW and inducing superconductivity in a fashion similar to chemical doping[3]. Therefore, unveiling the interplay among the lattice structure, magnetism and superconductivity seems to be crucial for an understanding of the nature of high-temperature superconductivity in pnictides.

Femtosecond pump-probe techniques provide new opportunities to investigate various quantum degrees of freedom and their interactions in complex systems. Direct information about the interplay between single-particle electronic states, collective modes, magnetization and lattice structure may be obtained by monitoring the real-time dynamics following perturbation by ultrashort optical pulses[4-6,15]. In particular, time-resolved multi-terahertz (THz) techniques extending over the far- and mid-infrared (MIR) spectral windows enable resonant probing of single-particle and collective low-energy excitations with a resolution of a few tens of femtoseconds[15].

We present a study of the dynamics of the complex conductivity in $BaFe_2As_2$ over the frequency range from 10 THz to 26 THz (corresponding to photon energies between 41 meV and 110 meV) following photo-excitation with a 12 fs pump pulse in the near infrared (NIR) centered at 1.55 eV. Single crystals of $BaFe_2As_2$ exhibiting a SDW transition at $T_{SDW}$ = 120 K were grown from Sn flux[16]. Temperature-dependent equilibrium spectra were measured by spectroscopic ellipsometry in the range from 12 meV to 500 meV and for temperatures between 10 K and 300 K[17]. Photo-induced changes in the complex conductivity were obtained by measuring reflection changes with few-cycle MIR pulses directly in amplitude and phase of the electric field and as a function of the delay time $t_D$ after excitation[15]. The spot size of the pump pulses was set to 110 µm and that of the probe pulses to 50 µm.

The equilibrium optical conductivity above and below $T_{SDW}$ is depicted in Fig. 1a by the red and blue lines, respectively. The opening of the SDW gap gives rise to a transfer of spectral weight from below to above the gap edge at 80 meV[18]. Fig. 1b shows the time evolution of the optical conductivity change induced by excitation with the NIR pump pulse at an absorbed fluence of $\Phi = 63$ µJ/cm$^2$ at a base temperature of the lattice of $T_L = 10$ K. After excitation, the optical conductivity rapidly increases below the crossing point and decreases above. The extrema are reached at $t_D = 180$ fs, where the optical conductivity agrees well with the equilibrium optical properties in the normal state (see green circles in Fig. 1a) suggesting that the SDW order has vanished. Fig. 1c presents the averaged responses in three distinct spectral regions below (A), around (B) and above (C) the gap edge, as indicated in Fig. 1b. The dynamics fits well to an exponential recovery with a time constant of 0.63 ps (black solid lines in Fig. 1c).

In addition to the exponential decay, weak temporal oscillations with a frequency of 5.5 THz become discernible in the reflectivity changes at higher excitation densities (see Fig. S3 in Supplementary Information). This frequency matches the one of an $A_{1g}$ mode which involves the c-axis displacement of arsenic with respect to the Fe square lattice[19] (see Fig. 3a). Coherent driving of this phonon mode has been observed in time-resolved optical reflection[20] and photoemission studies[21]. When exciting below $T_{SDW}$, the coherent oscillations appear as small modulations on top of the strong electronic signal due to the collapse and recovery of the SDW gap (see Fig. S2 in Supplementary Information). Notably, we find that these oscillations become a prominent feature when pumping above $T_{SDW}$. Fig. 2a shows the temporal changes of the MIR optical conductivity in the normal state recorded at $T_L = 134$ K and $\Phi = 530$ µJ/cm$^2$. Instead of the strong dispersive signal due to the gap closure in the SDW state, the conductivity now exhibits an overall increase and strong modulations of the optical conductivity with the $A_{1g}$ phonon frequency at 5.5 THz. Most importantly, we find that these oscillations have a very characteristic, dispersive behavior. As demonstrated in Fig. 2b, the oscillations in spectral regions A and C (red and blue circles, respectively) are out of phase with each other while the oscillatory response is strongly suppressed in region B (green circles). A quantitative analysis of the oscillatory component reveals that it matches closely the change in the optical conductivity that is induced in the equilibrium state by the SDW: the difference between the stationary conductivities in the normal state at 140 K and in the SDW state at 100 K is shown as a black line in Fig. 2c. Except for a scaling factor of 2.5, the spectral change in conductivity derived by taking the difference between the transient oscillation maxima and minima averaged over delay times $t_D$ between 280 fs and 1.12 ps closely follows this graph (see red circles in Fig. 2c). This congruence between the spectral shapes of the oscillatory response and the gap induced by the SDW represents the most important and surprising observation of our study. It indicates that even in the normal state well above $T_{SDW}$ the coherently driven $A_{1g}$ phonon periodically induces the SDW. This finding suggests that the macroscopic order parameter underlying the SDW exhibits a fast response to the collective nuclear motion with a modulation frequency as high as 5.5 THz.

In the following, we will develop a qualitative understanding of the coherent induction of SDWs in the normal state of $BaFe_2As_2$ which is supported by further experimental evidence. As discussed above, the ground state of pnictides may be tuned via external pressure and giant magneto-elastic coupling has been pointed out[3,8,22]. The distance of As from the Fe square lattice or the Fe-As-Fe bonding angle α marked in Fig. 3a is suggested to represent the most important structural parameter[3,9,23]. It was demonstrated that the maximum superconducting transition temperature is reached when the $FeAs_4$ unit forms an ideal tetrahedron with α = 109.47 degrees[3,9]. Since the $A_{1g}$ vibration of the arsenic ions modulates precisely this angle α, it might sensitively affect the ground state of the system. It was argued that due to the large polarizability of arsenic ions a small displacement can result in a significant modification of the band structure around the Fermi level and magnetic ordering further enhances this effect[24]. This prediction is supported by the strong modulations of the density of states around the Fermi level due to the $A_{1g}$ phonon, as observed in time and angle resolved photoemission experiments[21].

The Fermi surface of pnictides is composed of electron and hole pockets centered around the X and Γ points of the Brillouin zone, respectively[8,25]. When the sizes of the electron and hole pockets are comparable, large portions of the Fermi surface may be connected with a single scattering vector. This nesting of the Fermi surface is illustrated in Fig. 3b and may result in an opening of an energy gap[26]. Although the microscopic origin of magnetism in pnictides is still debated, it is widely assumed that the peculiar band structure intimately involves the Fermi surface nesting in forming the magnetic state[5]. In any case, when the $A_{1g}$ phonon modulates the band structure near the Fermi level, the nesting condition between the two bands will also be modified as depicted in Fig. 3b. Therefore, the coherent arsenic vibration could result in modulations not only of the density of states of the relevant bands but also of the SDW instability itself.

In our model, the effect of coherent lattice vibrations on the order parameter is expected to be strong even in the normal state near $T_{SDW}$ where it is obviously able to turn the SDW order on and off. Fig. 4a presents the temperature dependence of the modulation amplitude of the MIR conductivity due to the $A_{1g}$ phonon. It exhibits a broad maximum around $T_{SDW}$ followed by a pronounced decrease with increasing temperature. However, the amplitude of the coherent lattice vibrations depends on the Raman susceptibility for the near-infrared pump light. Hence it is expected to stay constant without a drastic change in the electronic structure[27]. This fact is well reflected by the temperature dependence of the oscillatory component of the reflectivity recorded at a probe photon energy of 1.55 eV, i.e. far above the SDW gap: The data show only a slight increase towards higher temperature in the normal state after a weak anomaly near $T_{SDW}$ (Fig. 4b). The opposite temperature dependences of the modulation amplitudes of MIR conductivity and NIR reflectivity support our assignment of the dispersive oscillatory component in the MIR range to a photo-induced ordering of SDWs: Despite the fact that the amplitude of the coherent lattice motion does not decrease with increasing temperature it becomes exceedingly difficult to induce the SDW order at $T \gg T_{SDW}$.

In order to test our model, it is instructive to compare the structural changes described by the phonon vibrations with those induced by temperature or external pressure: The absolute amplitude of the coherent lattice vibrations may be estimated based on the modulation of the NIR reflectivity and equilibrium dielectric constants (see Ref. 27 and Supplementary Information). We retrieve an amplitude of the As displacement as large as 0.8 pm for $\Phi = 530$ µJ/cm$^2$ at 134 K. This value corresponds to a modulation of the angle $\alpha$ by as much as 0.3 degree. It is reported that external pressure reduces $\alpha$ by 0.4 degree/GPa, where 1.5 GPa already suppresses the structural phase transition[3]. Correspondingly, the lattice vibration we induce is equivalent to an alternating internal pressure in the range of GPa. While the decrease of $\alpha$ by external pressure suppresses the SDW, the As displacement toward a larger angle might enhance ordering. The angle indeed increases by 0.1 to 0.5 degrees when BaFe$_2$As$_2$ is cooled down from the normal state to the SDW state[28,29]. This finding indicates that the amplitudes of the coherent phonon vibrations resulting from photo-excitation provide a deformation potential sufficient for the ultrafast ordering phenomena we observe.

A spin density wave order can be established only when the energy of the participating electrons remains smaller than the SDW gap energy. When the pump energy is delivered to the sample, the electrons are prepared in a non-equilibrium energy distribution within a few tens of femtoseconds. If we assume that all the absorbed energy is initially deposited in the electronic subsystem and efficient electron-electron scattering leads to a rapid thermalization within approximately 100 fs, then the system can be approximately described by an electronic temperature $T_e$ (as e.g. in a two temperature model[20]). The fluence of 530 µJ/cm² used in the measurement of Fig. 2 could increase $T_e$ up to 1100 K and prevent the formation of the SDW[20]. Ultrafast energy relaxation of normal-state electrons with a time constant shorter than 0.2 ps, however, reduces $T_e$ to approximately 500 K within one cycle of the $A_{1g}$ phonon oscillation[30]. As a result, the electronic energy $k_B T_e$ rapidly becomes much smaller than the modulated SDW gap energy of $2\Delta^m_{SDW} \approx 76$ meV ($2\Delta^m_{SDW}/k_B \approx 880$ K) observed in Fig. 2. It is noteworthy that while the static SDW state has been observed only in the orthorhombic structure, electronic anisotropy as well as magnetic fluctuations survive in the tetragonal structure up to temperatures well above the transition to long-range macroscopic order[25,31,32]. The phonon vibrations may drive the fluctuation coherently such that the transient order could develop at a higher temperature. Note that despite the ultrafast formation we find for the periodic buildup of the SDW, it is unlikely that this process will proceed instantaneously and some retardation should be expected. Such an inherent time scale may prevent the full development of SDW order and account for the reduction of the spectral weight in the transient case found in Fig. 2c. Further studies on the ultrafast structural dynamics of the system are desired to reveal the detailed interplay between the lattice and SDW order.

In conclusion, our observation of a coherent lattice vibration directly modulating a macroscopic order parameter on ultrafast time scales represents a qualitatively new form of coherent control of complex solids. Also, we demonstrate a crucial role of the lattice for the appearance of a SDW state in $BaFe_2As_2$. The remarkable fact that the magnetic order quasi-adiabatically follows fast lattice motion at a frequency as high as 5.5 THz is compatible with the scenario of a transient SDW order driven via nesting of the Fermi surface. Finally, despite the fact that band structure calculations[33] as well as some experiments[20,21] have suggested a weak electron-phonon coupling in pnictides, we observe a strong influence of collective nuclear motion on spin ordering. Therefore, an electron-lattice interaction mediated via spin-phonon coupling[22,24,34] may play an important role in the microscopic mechanism of the formation of SDW and superconducting states in the pnictides.

**Supplementary Information** is linked to the online version of the paper at www.nature.com/nature.

**Acknowledgements** We acknowledge financial support by the Schweizer Nationalfonds (SNF) under Grants No. PA00P2_129091 and No. 200020-129484, and by Deutsche Forschungsgemeinschaft via the Emmy Noether Program and SPP 1458, and by the Alexander von Humboldt Foundation.

**Author Contributions** K.W.K., J.D., R.H., A.L. planned the project; K.W.K. performed ellipsometry measurements; K.W.K., A.P., M.P. performed THz measurements; H.S., M.B. performed NIR/VIS measurements; K.W.K., A.P., H.S., M.B. analyzed data; T.W.

grew samples; and K.W.K., A.P., C.B., J.D., R.H., A.L. wrote the paper. All authors contributed to discussions and gave comments on the manuscript.

**Author Information** The authors declare no competing financial interests. Correspondence and requests for materials should be addressed to K.W.K. (kyungwan.kim@gmail.com) and R.H. (rupert.huber@physik.uni-regensburg.de).

**Figure 1 | Photo-induced transient changes of the mid-infrared conductivity in the spin density wave state. a**, Stationary optical conductivity in the spin density waves state at 10 K (blue line) and in the normal state at 140 K (red line). The conductivity of the excited state at a delay time of 180 fs after photo-excitation at 10 K is marked by green circles. **b**, Two-dimensional color map of the photo-induced change of the optical conductivity versus photon energy and delay time at $T_L$ = 10 K and $\Phi$ = 63 µJ/cm². **c**, Conductivity changes spectrally integrated over regions A, B and C as shown in **b**. The black solid lines represent fits to the data with a single exponential decay time $\tau$ = 0.63 ps and a small offset.

**Figure 2 | Photo-induced transient changes of the MIR conductivity in the normal state slightly above $T_{SDW}$. a,** Two-dimensional map of the photo-induced change of the optical conductivity at $T_L$ = 134 K and $\Phi$ = 530 µJ/cm² and **b,** its averages in spectral regions A, B and C. Spectra in regions A and B are vertically shifted for clarity. Red dotted lines at maxima/minima emphasize the out-of-phase oscillations in different frequency regions. **c**, Comparison of the dispersion of the oscillatory signal $\Delta\sigma_1^{osc}(\omega)$ with the one of the SDW induced changes in the equilibrium state $\Delta\sigma_1^{eq}(\omega) = \sigma_1^{100K}(\omega) - \sigma_1^{140K}(\omega)$. $\Delta\sigma_1^{osc}(\omega)$ is obtained from the difference spectrum between averages of maxima and minima in the interval of time delays between 0.28 ps and 1.12 ps after subtracting an exponential decay component of the electronic signal.

**Figure 3 | Schematic visualization of the $A_{1g}$ vibration and its effect on the band structure**. **a**, Real space sketch of the eigenvector of the $A_{1g}$ mode at 5.5 THz which

involves the motion of arsenic ions with respect to the Fe square lattice. The Fe-As-Fe angle α is modulated depending on the position of arsenic ions. **b**, Reciprocal space sketch of the electron and hole bands at the X and Γ points of the Brillouin zone connected by the antiferromagnetic nesting vector $q_{AF}$. The effect of the arsenic displacement on the band structure is depicted with dotted lines. Blue and red lines correspond to cases of arsenic displacements along blue and red arrows in **a**, respectively[24]. Via the effect of Fermi surface nesting, the modulations of the electronic bands can influence also the spin density wave instability.

**Figure 4 | Temperature dependence of the relative modulation amplitude due to the coherent phonon at 5.5 THz. a**, Red squares show modulation amplitudes obtained from $\Delta\sigma_1$ (averaged over the range from 41 meV to 62 meV). Data obtained from one-dimensional measurements are depicted as black circles and correspond to a spectral integration over the entire bandwidth of the multi-THz probe between 41 meV and 110 meV). All data are recorded at an excitation density of $\Phi = 530$ μJ/cm². Absolute values are scaled for comparison. **b**, Modulation amplitudes of reflectivity at a probe photon energy of 1.55 eV and at an excitation density of $\Phi = 200$ μJ/cm². Amplitudes are obtained from fits to a sinusoidal function and the error bars indicate 95% confidence intervals for the fitting parameter (see Supplementary Information).

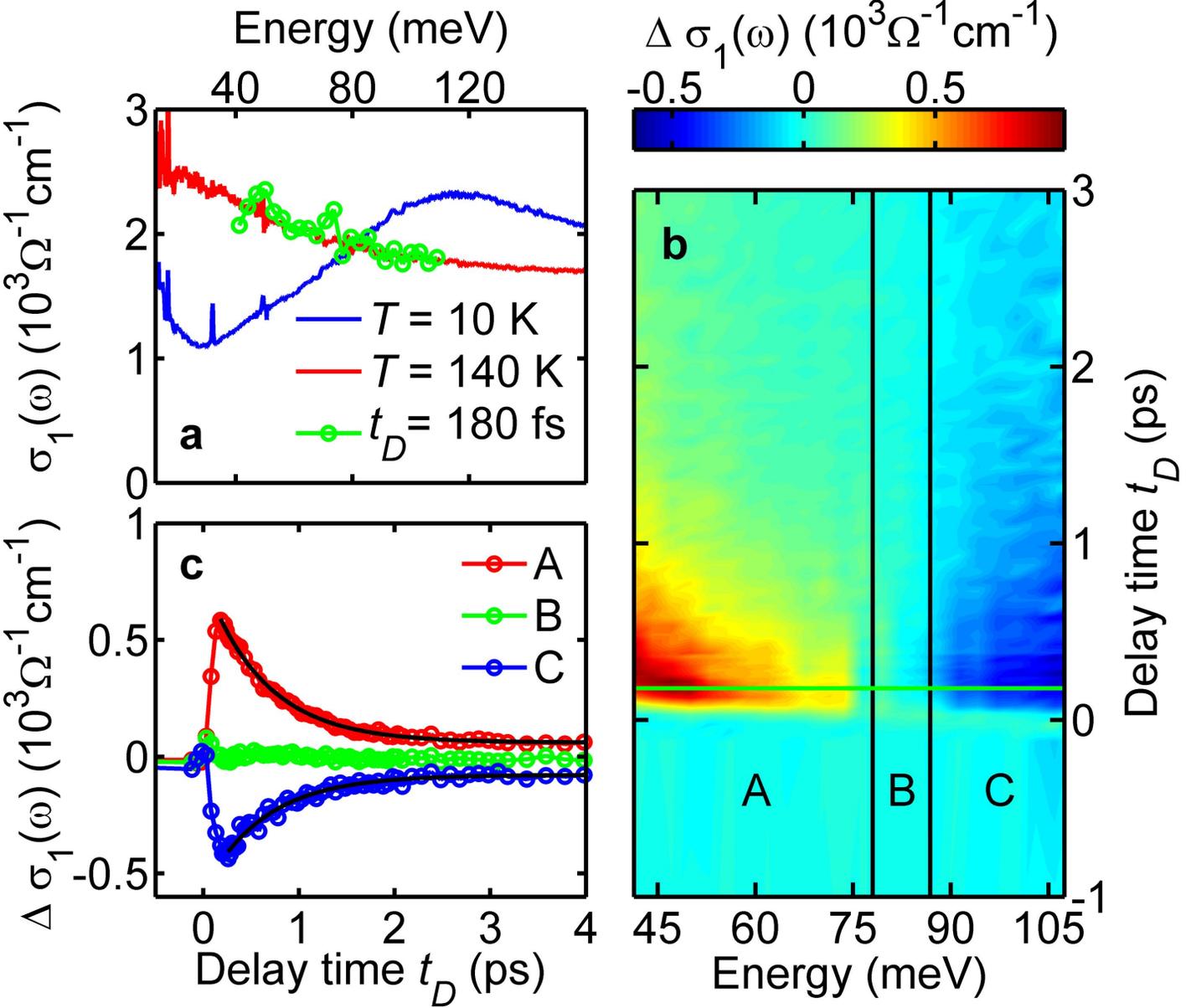

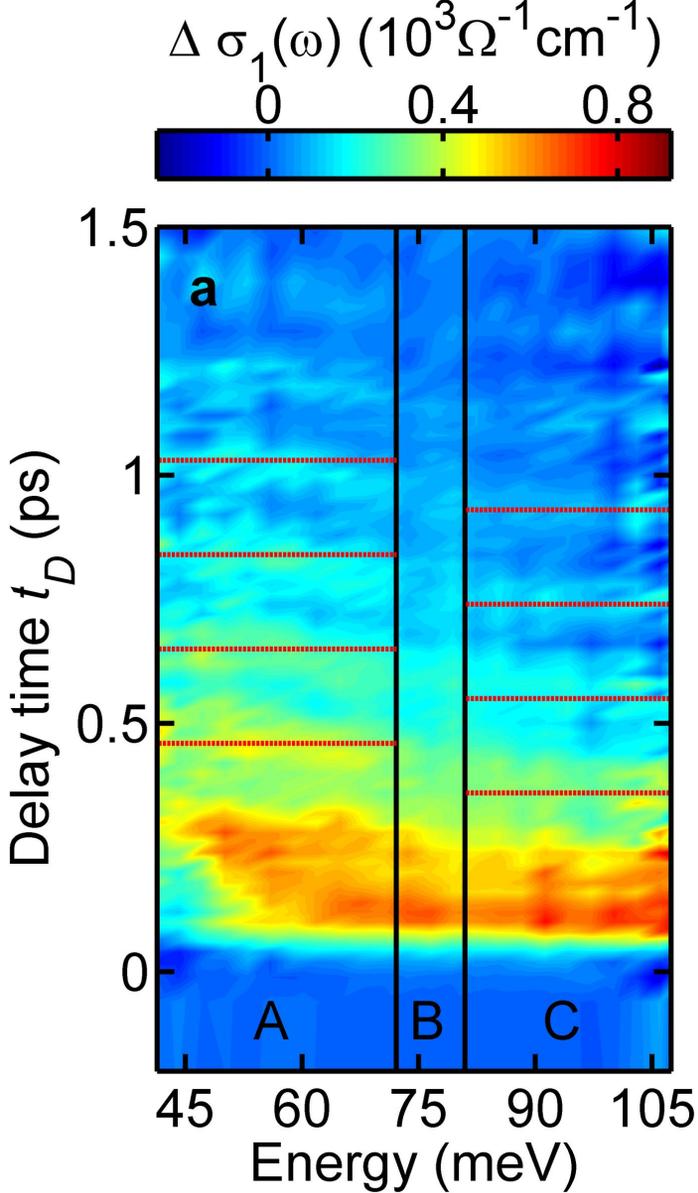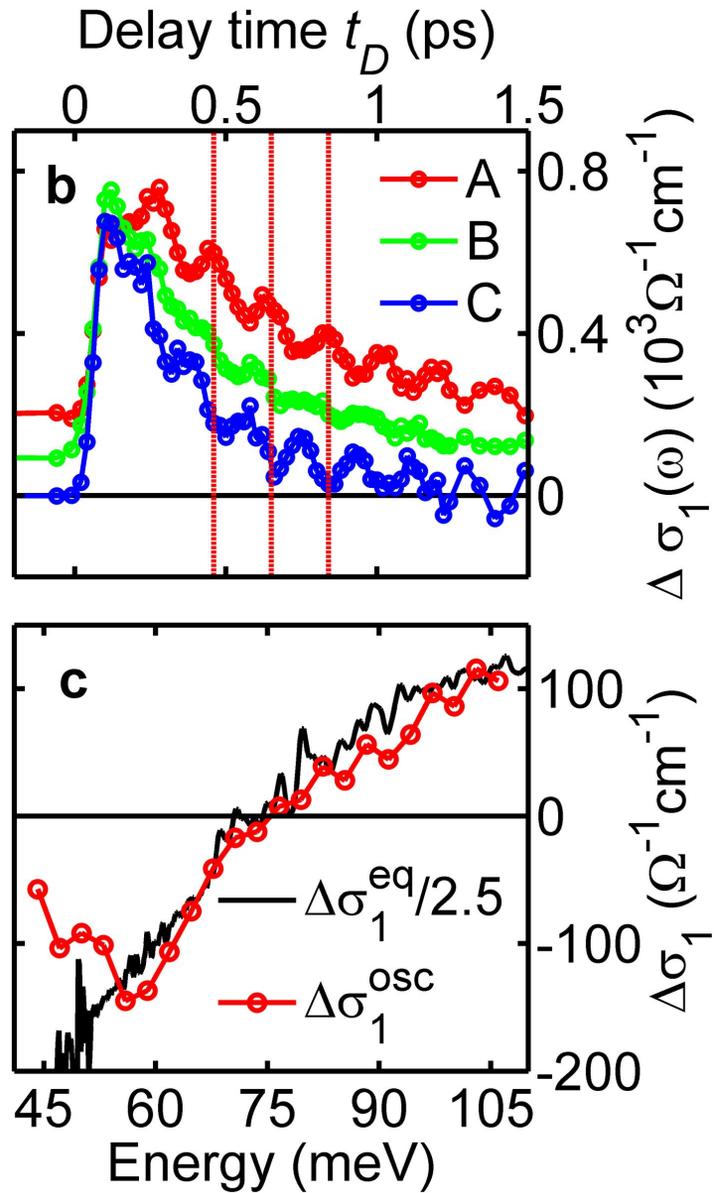

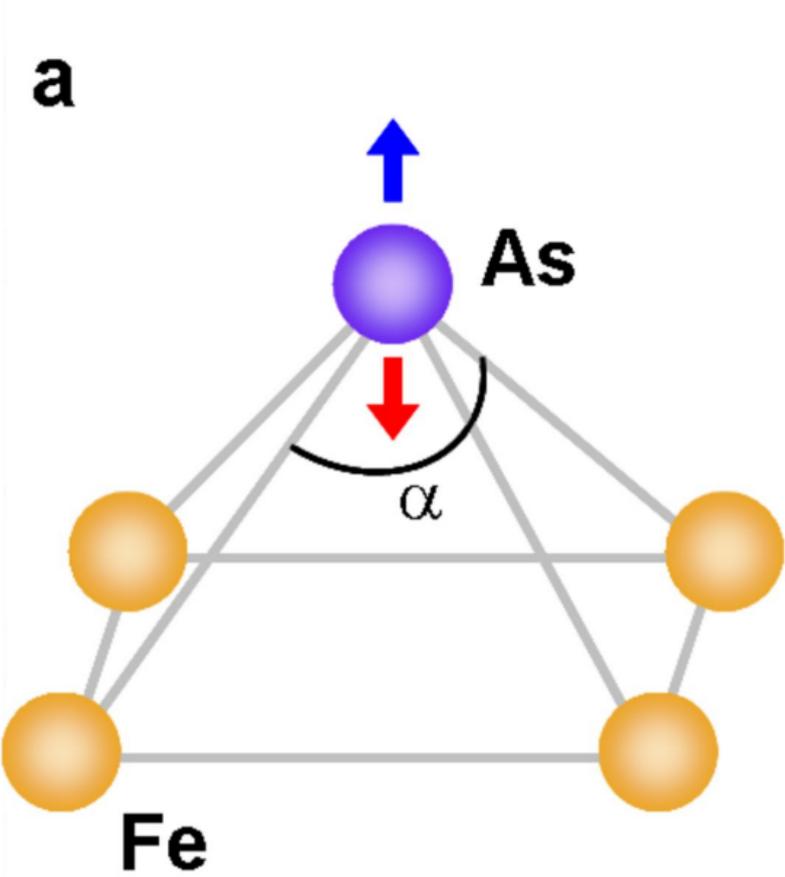 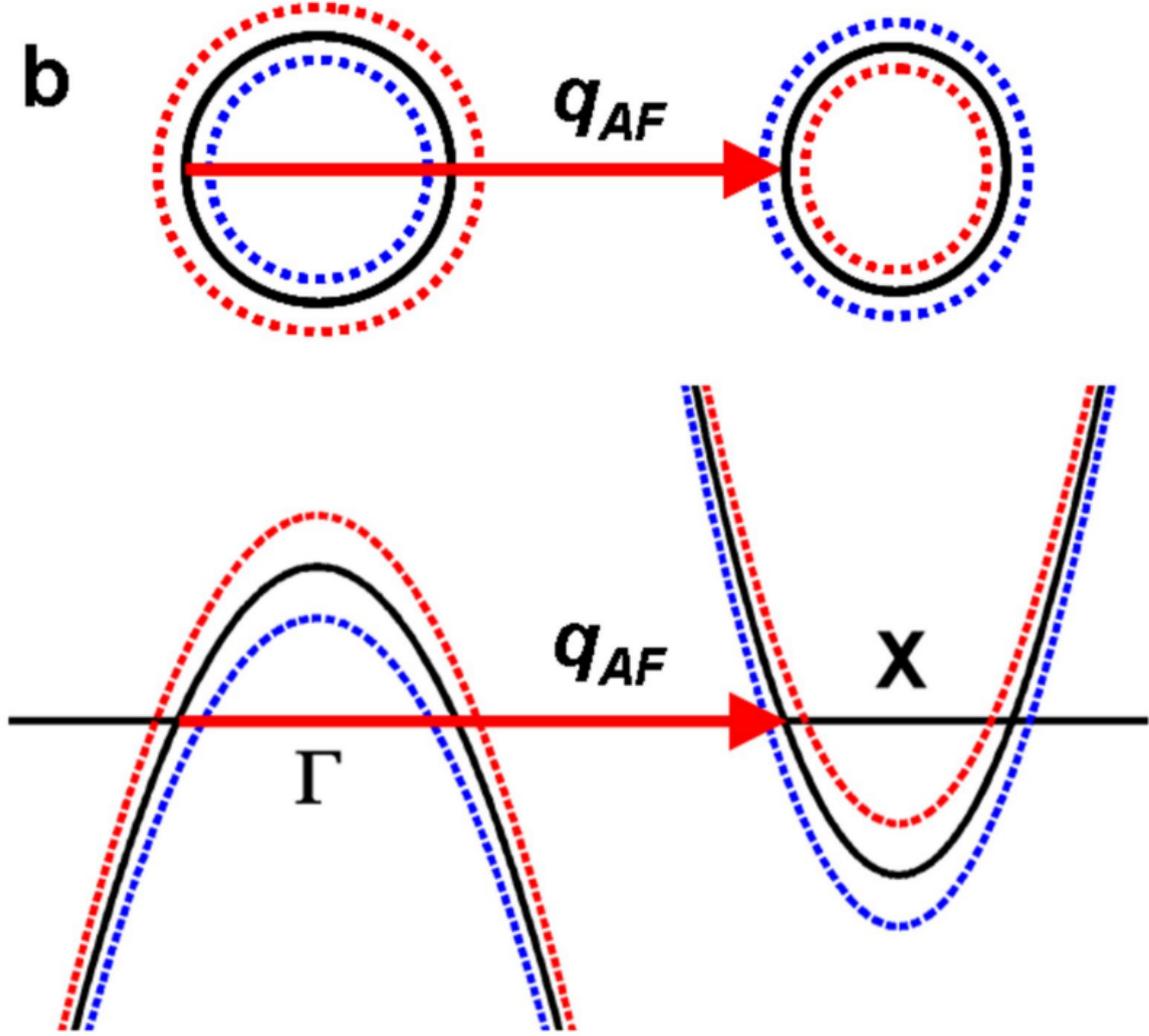